\documentclass[prl,aps,showpacs,reprint,superscriptaddress,amsmath,amssymb,floatfix,10pt]{revtex4-1} 
\usepackage{amsmath,amssymb,graphicx}
\usepackage[colorlinks=true, citecolor=blue, linkcolor=blue, urlcolor=blue]{hyperref}
\usepackage{array}

\begin{document}

\title{Superresolving Imaging of Irregular Arrays of Thermal Light Sources using Multiphoton Interferences}

\author{Anton Classen}
	\affiliation{Institut f\"ur Optik, Information und Photonik, Universit\"at Erlangen-N\"urnberg,
	91058 Erlangen, Germany}
	\affiliation{Erlangen Graduate School in Advanced Optical Technologies (SAOT), Universit\"at Erlangen-N\"urnberg,
	91052 Erlangen, Germany}
	
	\author{Felix Waldmann}
	\affiliation{Institut f\"ur Optik, Information und Photonik, Universit\"at Erlangen-N\"urnberg,
	91058 Erlangen, Germany}
	
\author{Sebastian Giebel}
	\affiliation{Institut f\"ur Optik, Information und Photonik, Universit\"at Erlangen-N\"urnberg,
	91058 Erlangen, Germany}
	
\author{Raimund Schneider}
	\affiliation{Institut f\"ur Optik, Information und Photonik, Universit\"at Erlangen-N\"urnberg,
	91058 Erlangen, Germany}
	\affiliation{Erlangen Graduate School in Advanced Optical Technologies (SAOT), Universit\"at Erlangen-N\"urnberg,
	91052 Erlangen, Germany}

\author{Daniel Bhatti}
	\affiliation{Institut f\"ur Optik, Information und Photonik, Universit\"at Erlangen-N\"urnberg,
	91058 Erlangen, Germany}
	\affiliation{Erlangen Graduate School in Advanced Optical Technologies (SAOT), Universit\"at Erlangen-N\"urnberg,
	91052 Erlangen, Germany}

\author{Thomas Mehringer}
	\affiliation{Institut f\"ur Optik, Information und Photonik, Universit\"at Erlangen-N\"urnberg,
	91058 Erlangen, Germany}
	\affiliation{Erlangen Graduate School in Advanced Optical Technologies (SAOT), Universit\"at Erlangen-N\"urnberg,
	91052 Erlangen, Germany}		
		
\author{Joachim von Zanthier}
	\affiliation{Institut f\"ur Optik, Information und Photonik, Universit\"at Erlangen-N\"urnberg,
	91058 Erlangen, Germany}
	\affiliation{Erlangen Graduate School in Advanced Optical Technologies (SAOT), Universit\"at Erlangen-N\"urnberg,
	91052 Erlangen, Germany}	

\begin{abstract}
We propose to use multiphoton interferences of photons emitted from statistically independent thermal light sources in combination with linear optical detection techniques to reconstruct, i.e., image, arbitrary source geometries in one dimension with subclassical resolution. The scheme is an extension of earlier work [Phys. Rev. Lett. 109, 233603 (2012)] where $N$ regularly spaced sources in one dimension were imaged by use of the $N$th-order intensity correlation function. Here, we generalize the scheme to reconstruct any number of independent thermal light sources at arbitrary separations in one dimension exploiting intensity correlation functions of order $m \geq 3$. We present experimental results confirming the imaging protocol and provide a rigorous mathematical proof for the obtained subclassical resolution. 
\end{abstract}

\maketitle

Higher order interferences with photons emitted by statistically independent light sources are an active field of research with the potential to increase the resolution in spectroscopy, lithography and interferometry \cite{Wineland2004a,Kok2000a,Shih2001a,Steinberg2004a,Zeilinger2004a,Scully2006}, as well as in imaging and microscopy \cite{Teich1997,Muthukrishnan2004,Ariunbold2004,Thiel2007,Shapiro2008a,Lloyd2008a,Shapiro2009,Oppel2012a,Genovese2013a,Genovese2014a,Pearce2015}. 
So far, subclassical resolution has been achieved by using entangled photons \cite{Shih2001a,Muthukrishnan2004}, but it was also shown that initially uncorrelated light fields - non-classical as well as classical - can be employed for that purpose \cite{Shapiro2009,Oppel2012a,Genovese2013a,Genovese2014a,Pearce2015}.
Recently, Oppel \textit{et al.} presented a detection scheme that allows to determine the source distance $d$ for an array of $N$ equidistant thermal light sources (TLS) with subclassical resolution by measuring the $N$th-order spatial intensity correlation function \cite{Oppel2012a}.

Here, we show that the scheme presented in \cite{Oppel2012a} can be generalized to reconstruct, i.e., image, any number of independent TLS at arbitrary separations in one dimension by exploiting photon correlation functions of order $m \geq 3$. 
Measuring higher order correlations enables to isolate the spatial frequencies of the setup allowing to determine the source distribution with a resolution below the classical Abbe limit. We  outline the imaging protocol and present experimental results verifying the theoretical predictions. A physical explanation and rigorous mathematical proof of the protocol and the spatial frequency filtering process is given in the Supplemental Material.

\begin{figure}[b]
\centering
\includegraphics[width=0.4\textwidth]{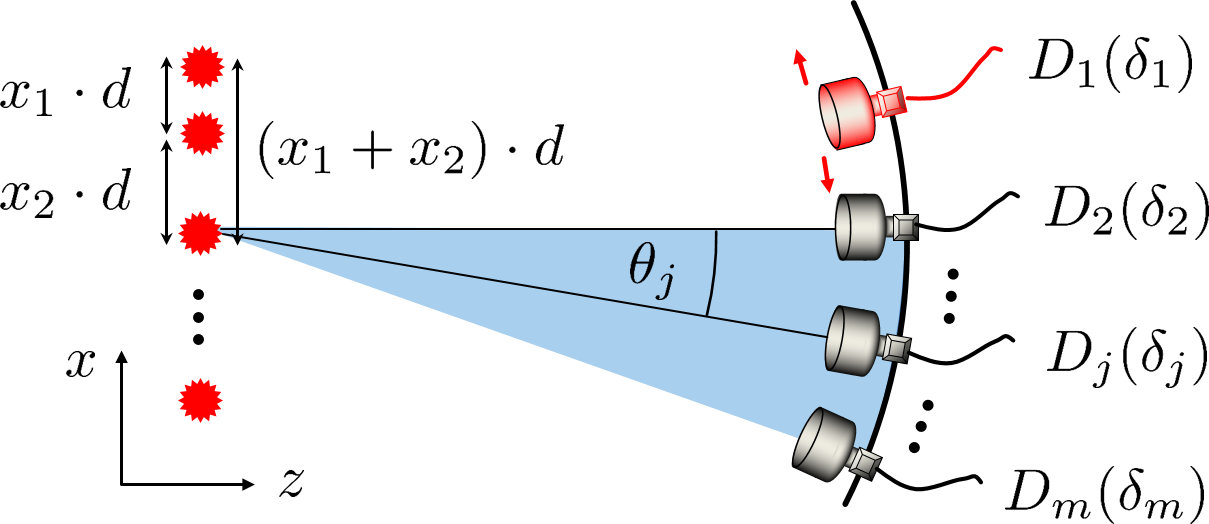}
\caption{(color online) Scheme of the considered setup: $N$ irregularly arranged TLS are aligned on a grid in one dimension with lattice constant $d$ such that $|\mathbf{R}_{l+1} - \mathbf{R}_{l}| = x_l \cdot d$, with $x_l \in \mathbb{N}$, $l=1,\ldots, N-1$. In the far field of the sources $m$ detectors $D_j$, $j = 1, \ldots, m$ measure the intensities at $\mathbf{r}_1, \ldots, \mathbf{r}_m$, with $ \delta_j = \delta_j (\mathbf{r}_j)  = k d \sin \left[\theta_j (\mathbf{r}_j) \right] $.}
	\label{fig:inv_setup}
\end{figure}

We assume $N$ TLS aligned on a grid in one dimension with lattice constant $d$ at arbitrary separations, such that $|\mathbf{R}_{l+1} - \mathbf{R}_{l}| = x_l \cdot d$, with $x_l \in \mathbb{N}$, $l=1,\ldots, N-1$. The source geometry is thus determined by the lattice constant $d$ and the $N-1$ adjacent source distances $\mathbf{x} = (x_1,x_2, \ldots, x_{N-1})$,
whereas the spatial frequencies of the system are given by the tuple of source pair distances $\{  \mathbf{\xi} \} \equiv \{ (x_1); \, (x_1+x_2); \ldots ; \, (x_{l_1}+\cdots+x_{l_2}) ; \ldots ; \, (x_1+\cdots+x_{N-1})  \}$ (see Fig.~\ref{fig:inv_setup}).

To access the set of spatial frequencies $\{  \mathbf{\xi} \}$ we make use of the normalized spatial $m$th-order intensity correlation function $g^{(m)}_{N}(\mathbf{r}_1,\ldots,\mathbf{r}_m)$ obtained by correlating the intensities at positions $\mathbf{r}_1, \ldots, \mathbf{r}_m$ 
in the far field \cite{Glauber1963} 
\begin{equation}
g^{(m)}_{N}(\mathbf{r}_1,\ldots,\mathbf{r}_m) \equiv \frac{\langle :  \prod_ {j=1}^{m}  \hat{E}^{(-)}(\mathbf{r}_j) \hat{E}^{(+)}(\mathbf{r}_j)  : \rangle_{\rho} }{\prod_ {j=1}^{m} \langle  \hat{E}^{(-)}(\mathbf{r}_j) \hat{E}^{(+)}(\mathbf{r}_j) \rangle_{\rho}} \, .
\label{eq:3}
\end{equation}
Here, $\langle : \, \cdot \, : \rangle_{\rho}$ denotes the (normally ordered) quantum mechanical expectation value for a system in the state $\rho$ and $\hat{E}^{(-)}(\mathbf{r}_j)$ and $\hat{E}^{(+)}(\mathbf{r}_j)$ are the positive and negative frequency parts of the total electric field operator at position $\mathbf{r}_j$,
given by $\hat{E}^{(+)}(\mathbf{r}_j) = \left[ \hat{E}^{(-)}(\mathbf{r}_j) \right]^{\dagger} \propto \sum_{l} e^{i k r_{l j}} \hat{a}_l$ \cite{Oppel2012a}. In the last expression, $\hat{a}_l$ is the annihilation operator of a photon emitted by source $l$ at $\mathbf{R}_l$ and $r_{l j} = |\mathbf{R}_l - \mathbf{r}_j | $. Note that since we assume the emitters to be statistically independent, the state of the field is given by $\rho = \otimes_l \rho_l$, with $\rho_l = \sum_n P_l(n) \left| n \right> \langle n |$, where $P_l(n)$ is the photon number distribution of source $l$ \cite{Bhatti2016}.

In the case of a regular source arrangement with $N$ equidistant TLS at separation $d$
and $m-1$ detectors placed at $\mathbf{r}_2 = \cdots = \mathbf{r}_m = 0$  the $m$th-order correlation function as a function of the position of the first detector takes the form $g^{(m)}_N(\mathbf{r}_1; 0) \equiv g^{(m)}_N(\delta_1; 0) \propto \text{c} + \sum_{l=1}^{N-1} (N-l) \cos (l \delta_1)$, with  $ \delta_j = \delta_j (\mathbf{r}_j)  = k d \sin \left[ \theta_j (\mathbf{r}_j) \right] $ \cite{Oppel2014,Bhatti2016}. Note that  $g^{(m)}_N(\mathbf{r}_1)$ displays \textit{all} $N-1$ different spatial frequencies $l d$, $l = 1, \ldots, N-1$, of the setup, equally obtained when measuring the intensity distribution of a coherently illuminated $N$ slit grating with slit separation $d$.

For an irregular source arrangement with arbitrary separations 
it turns out that by placing $m-1$ detectors at the so-called \textit{magic positions} \cite{Oppel2012a}
\begin{equation}
\delta_j = 2\pi (j-2)/ (m-1) \qquad j=2,\ldots,m \, ,
\label{eq:4}
\end{equation}
all spatial frequencies of the source arrangement are suppressed in $g^{(m)}_{N}(\delta_1)$,
except those fulfilling the condition 
\begin{equation}
  \kappa \, (m-1) = (x_{l_1} + \cdots + x_{l_2}) \in   \{ \mathbf{\xi} \}   \; ,
\label{eq:6}
\end{equation}
with  $\kappa \in \mathbb{N}$.  In this case the $m$th order intensity correlation function takes the form \cite{SupplementalMaterial}
\begin{equation}
g^{(m)}_{N}(\delta_1)= A_0^{(m)} +\sum_{\kappa}{A_{\kappa}^{(m)} \, \cos(\kappa \, (m-1) \, \delta_1)} \ ,
\label{eq:5}
\end{equation}
where $A_\kappa^{(m)}$ is the amplitude of the modulation with frequency $\kappa \, (m-1)$; if  
no element of $\{  \mathbf{\xi} \}$ fulfills Eq.~(\ref{eq:6}),
i.e., all spatial frequencies $(x_{l_1}+ \cdots + x_{l_2}) \in  \{  \mathbf{\xi} \}$ differ from $\kappa (m-1) $, we obtain $g^{(m)}_{N}(\delta_1)= A_0^{(m)} = \text{const}$. 

Note that one can access the magic positions by changing the positions  $\mathbf{r}_2, \ldots, \mathbf{r}_N$ of the detectors $D_2, \ldots, D_N$ while monitoring the interference pattern $g^{(m)}_{N}(\delta_1)$ until a modulation of the form $\sum_{\kappa}{A_{\kappa}^{(m)} \, \cos(\kappa \, (m-1) \, \delta_1)}$
appears \footnote{If no modulation appears in $g^{(m)}_{N}(\delta_1)$ this means that the setup contains no spatial frequency fulfilling the condition of Eq.~(\ref{eq:6}).}. In this case the relative phase relation $\delta_j - \delta_{j-1} = 2 \pi/(m-1)$, $j = 3, \ldots, m$ is fulfilled (see Eq.~(\ref{eq:4})). The lattice constant $d$ can then be determined from $\delta_j$ and $\delta_{j-1}$ via $ d = \lambda / \{   (m-1) [\sin(\theta_j) -\sin(\theta_{j-1})]  \}$.
Note further that, in view of Eqs.~(\ref{eq:6}) and (\ref{eq:5}),
the regular source distribution discussed in \cite{Oppel2012a} is merely a special case of the outlined imaging protocol with $m=N$. Indeed, for $x_1=x_2=\cdots=x_{N-1} \equiv 1$ we obtain for $m=N$ 
\begin{equation}
g^{(N)}_{N\text{}}(\delta_1) = A_0^{(N)} + A^{(N)}_N \cos[(N-1)\delta_1]  \, .
\label{eq:6a}
\end{equation}
However, in contrast to Eq.~(\ref{eq:6a}), the spatial frequency filtering process of Eqs.~(\ref{eq:6}) and (\ref{eq:5}) neither depends on the number of sources, i.e., it can be applied for $m \neq N$, nor does it rely on a particular source geometry $\mathbf{x}$ \cite{SupplementalMaterial}.

Measuring $g^{(m)}_{N}(\delta_1)$ for $m \geq 3$ allows to determine all spatial frequencies $\in \{ \mathbf{\xi} \}$ fulfilling Eq.~(\ref{eq:6}).
However, since not all of the $N(N-1)/2$ spatial frequencies of the unknown source geometry $\mathbf{x}$ are necessarily different, the scheme  has access only to the smaller set of all different spatial frequencies
\begin{equation}
\mathbf{F} \equiv \{  \text{all different spatial frequencies}    \in  \{ \mathbf{\xi} \} \}  = \left\{ f_i \right\}  \, .
\label{eq:7}
\end{equation}
$\mathbf{F}$ still contains a large amount of information, narrowing down the set of possible source geometries substantially so that in most cases a unique solution can be obtained.

\begin{figure}[ht!]
  \centering
	\includegraphics[width=0.4\textwidth]{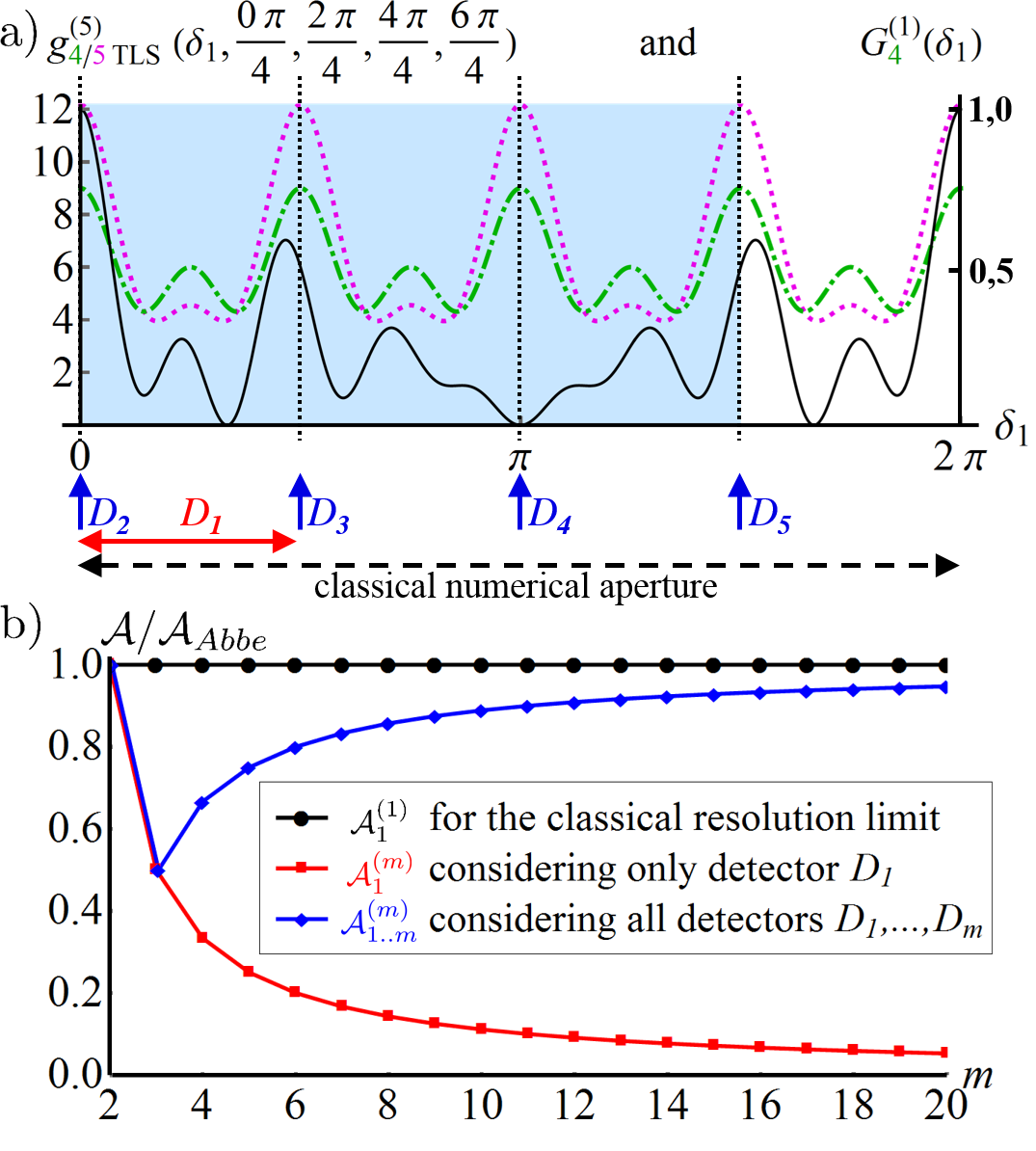}
\caption{(color online) a) $G^{(1)}_{4}(\delta_1)$ for $N=4$ coherently emitting sources with distances x=(3,1,4) (solid (black) curve) and $g^{(5)}(\delta_1)$ for both scenarios of Eq.~(\ref{eq:8}): $N=4$ TLS (dotted-dashed (green) curve) and $N=5$ TLS (dotted (magenta) curve); for the latter two cases four detectors are fixed at the magic positions. The numerical aperture  $\mathcal{A}^{(1)}_{1}$ required by the classical Abbe limit is indicated by the dashed (black) arrow below the x-axis; the numerical aperture $\mathcal{A}^{(m)}_{1}$ required for $D_1$ to scan from one to the next prinicipal maxima is indicated by the solid (red) arrow below the x-axis; the numerical aperture $\mathcal{A}^{(m)}_{1\ldots m}$ required by all detectors $D_j$, $j = 1, \ldots, 5$ is indicated by the blue shaded area (see Fig.~\ref{fig:inv_setup}). b) numerical apertures $\mathcal{A}^{(1)}_{1}$, $\mathcal{A}^{(m)}_{1}$ and $\mathcal{A}^{(m)}_{1\ldots m}$ (in units of $\mathcal{A}^{(1)}_{1} = \mathcal{A}_{\text{Abbe}}$) as a function of correlation order $m$. As can be seen, $\mathcal{A}^{(m)}_{1}$ and $\mathcal{A}^{(m)}_{1\ldots m}$ are always smaller than $\mathcal{A}^{(1)}_{1}$.}
  \label{fig:Amplitudes}
\end{figure}

Consider for example the case $\mathbf{x}=(3,1,4)$. Here, the set of different spatial frequencies is given by $\mathbf{F}= \{ 1,3,4,5,8 \}$. Measuring all intensity correlation functions of order $3 \leq m \leq 9$ leads to a unique solution for the number and distribution of sources 
\begin{equation}
\begin{aligned}
\textbf{F} = \{1,3,4,5,8\}   \rightarrow & \hspace{3.4mm}  N=4   \quad \text{with} \quad \mathbf{x}=(3,1,4) \; ,
\end{aligned}
\label{eq:7a}
\end{equation}
i.e., a unique reconstruction (imaging) of the unknown source geometry can be achieved. By contrast, for the set of spatial frequencies 
$\textbf{F} = \{1,3,4,5,8,9\}$ two possible solutions for the unknown source geometry exist, namely
\begin{equation}
\textbf{F} = \{1,3,4,5,8,9\} \rightarrow  \begin{cases}
   N=4   \quad  \mathbf{x}=(1,3,5) \\
   N=5   \quad  \mathbf{x}=(1,3,1,4)  
\end{cases}
\label{eq:8}
\end{equation}
To remove the remaining ambiguity, additional information can be extracted from the amplitudes $A_k^{(m)}$ of the correlation functions $g^{(m)}_{N}(\delta_1)$, $m \geq 3$ (cf. Eq.~(\ref{eq:5})). As an example, we display $g^{(5)}(\delta_1)$ for the two scenarios of Eq.~(\ref{eq:8}) in Fig.~\ref{fig:Amplitudes}. The difference in (relative) amplitudes is clearly visible enabling a discrimination between the two solutions, so that again an unambiguous reconstruction is obtained.

\begin{figure*}
\centerline{
\includegraphics[width=2.0\columnwidth]{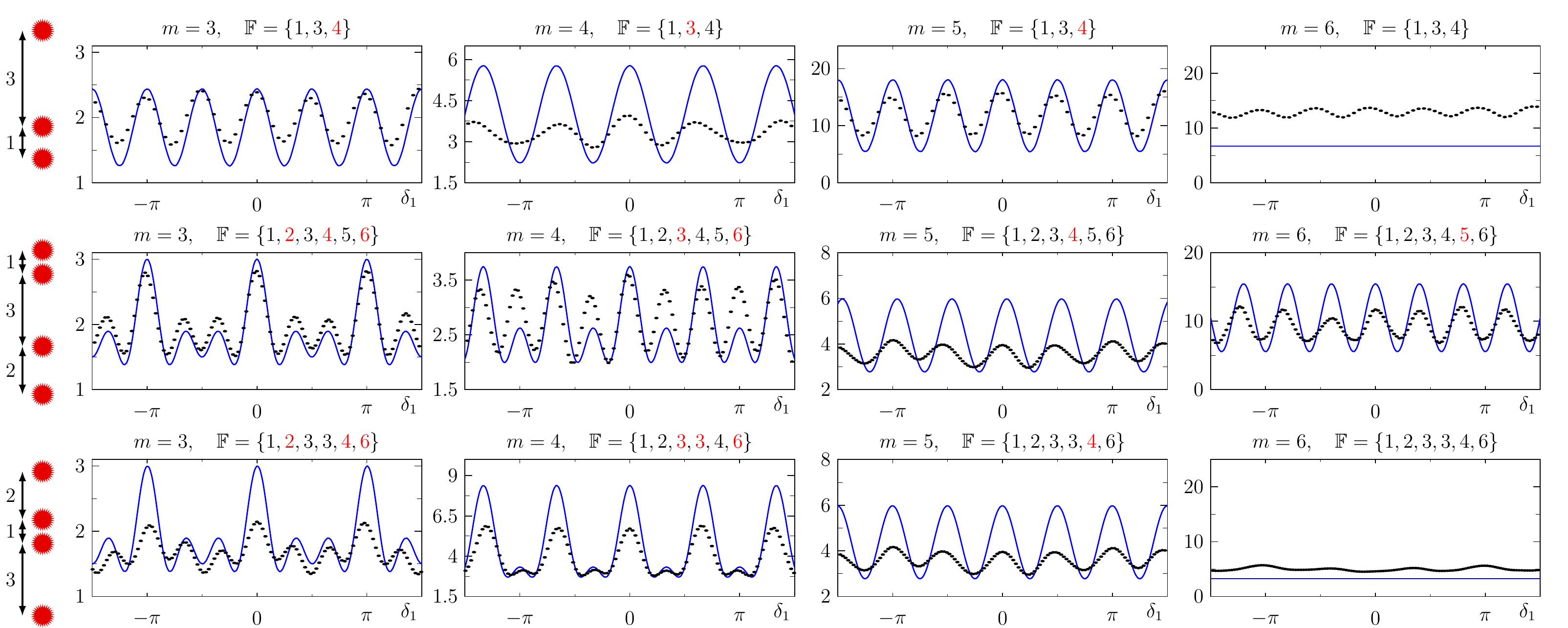}
}
\caption{(color online) Measured $m$th-order correlation function $g^{(m)}(\delta_1)$ for $m = 3, \ldots, 6$ (dotted  (black) curves), where $m-1$ detectors are placed at the magic positions (see Eq.~(\ref{eq:4})), together with the theoretically expected signals according to Eq.~(\ref{eq:5}) (solid (blue) curves) for the three source configurations shown on the left. The two lower configurations, having an equal set of source distances but different source arrangements and thus different set of spatial frequencies $\mathbf{F} \equiv  \left\{ f_i \right\}$, can be clearly distinguished by the imaging protocol.} 
	\label{fig:single_shots}
\end{figure*}
\begin{table*}
		\def\arraystretch{1}
	\begin{tabular}{| c ||c |c|c| c|c| c|c| c| }
  \hline
  $\mathbf{F}$ & \multicolumn{2}{c}{ $\mathbf{m=3}$}&\multicolumn{2}{c}{$\mathbf{m=4}$}&\multicolumn{2}{c|}{$\mathbf{m=5}$}&\multicolumn{2}{c|}{$\mathbf{m=6}$} \\
&$\overline{f_i}$ & $\overline{A_i^{(3)}}$ &$\overline{f_i}$ & $\overline{A_i^{(4)}}$ &$\overline{f_i}$ & $\overline{A_i^{(5)}}$
	&$\overline{f_i}$ & $\overline{A_i^{(6)}}$  \\
	\hline
$\left\{1,3,4\right\}$ &$ 4.02\pm0.01$ & $0.32\pm0.08$ & $2.93\pm0.03$ & $0.51\pm0.19$ & $4.02\pm0.01$ & $2.47\pm0.88$ & $3.90\pm0.31$ & $1.05\pm0.66$      \\
	\hline
&$1.96\pm0.01$ & $0.20\pm0.04$ & $3.02\pm0.02$ & $0.08\pm0.01$ & $3.98\pm0.01$ & $1.09\pm0.24$ & $4.93\pm0.01$ & $1.28\pm0.16$\\
$\left\{1,2,3,4,5,6\right\}$ &$3.98\pm0.01$ & $0.25\pm0.03$ & $5.94\pm0.01$ & $0.59\pm0.15$ & 					  & 					  & 					  & 				  \\
&$5.94\pm0.01$ & $0.33\pm0.05$ &					 	  & 					  &						  & 					  &							&	 							     \\
	\hline
&$2.09\pm0.01$ & $0.18\pm0.02$ & $3.01\pm0.01$ & $1.29\pm0.14$ & $4.00\pm0.01$ & $0.56\pm0.13$ & $2.99\pm0.55$ & $0.58\pm0.52$      \\
$\left\{1,2,3,4,6\right\}$&$3.97\pm0.01$ & $0.05\pm0.01$ & $6.06\pm0.01$ & $0.51\pm0.08$ &						  &  					  & 					  &    							  \\
&$6.06\pm0.01$ & $0.17\pm0.03$ & 						&							&							& 					  & 						&							       \\
	\hline
\end{tabular} 
\caption{Experimentally measured mean values for the spatial frequencies $\left\{f_i\right\}$ and corresponding amplitudes $A_i^{(m)}$, obtained for the correlation orders $m = 3, \ldots, 6$ according to Eqs.~(\ref{eq:6}) and (\ref{eq:5}), together with their standard deviations.}
	\label{tab:exp_results}
\end{table*}

Note that the determination of \textbf{F} makes small demands to the experimental data as only the spatial frequencies are to be identified (see Fig.~\ref{fig:single_shots}). The second step - reconstructing the source geometry $\mathbf{x}$ from $\mathbf{F}$ - sometimes requires a better data quality as in order to remove ambiguities the amplitudes $A_k^{(m)}$ of the modulations have to be taken into account.

The proposed imaging technique allows to reconstruct the source geometry $\mathbf{x}$ with a resolution below the Abbe limit. According to Abbe, for a given numerical aperture $\mathcal{A}$, the smallest resolvable distance is given by $d_{\text{min}}= \lambda / (2 \mathcal{A})$, 
where $\lambda$ is the wavelength of the light emitted by the sources \cite{BornWolf1999}. 
The range $\delta_1 \in [0 , 2 \pi]$ is required in the far field to resolve this distance as two adjacent principal maxima are separated by $\Delta \delta_1 =  2 \pi$ (see dashed (black) arrow in Fig.~\ref{fig:Amplitudes}a). By contrast, using the imaging protocol outlined above, the moving detector $D_1$ requires only the range $\Delta \delta_1 =  2 \pi/(m-1)$ to scan two adjacent maxima, as the fringe spacing is reduced by $(m-1)$ (see Eq.~(\ref{eq:5})). Due to the reduced numerical aperture $\mathcal{A}^{(m)}_{1}$ required for the moving detector, the resolution for the moving detector is enhanced by the same factor, i.e., overcoming the classical resolution limit by $(m-1)$ (see (red) squares in Fig.~\ref{fig:Amplitudes}b) \cite{Oppel2012a}. Considering the angular range of \textit{all} detectors, i.e., including the detectors placed at the magic positions, the required numerical aperture $\mathcal{A}^{(m)}_{1 \ldots m}$ increases (see (blue) diamonds  in Fig.~\ref{fig:Amplitudes}b). However, $\mathcal{A}^{(m)}_{1 \ldots m}$ remains below the aperture $\mathcal{A}^{(1)}_{1}$ required by the Abbe limit for all $m \geq 3$ (see Fig.~\ref{fig:Amplitudes}b).
The proposed imaging protocol is thus able to reconstruct the source geometry with subclassical resolution. Moreover, it allows to determine the spatial frequencies of the source ensemble with a substantially reduced number of fit parameters in comparison to classical imaging techniques. In the former case only one or at most few spatial frequencies have to be determined from $g^{(m)}(\delta_1)$, whereas in the case of classical imaging techniques \textit{all} spatial frequencies have to be identified in the Fourier plane at once.

For an experimental demonstration of the proposed imaging technique we used up to four statistically independent pseudothermal light sources (see Fig.~\ref{fig:single_shots}). The pseudo-TLS were realized by use of a He-Ne laser at $\lambda=$ $632.8$~nm coupled into multimode fibers of diameter $\sim 50\,\mu$m. 
The superposition of many modes in a given multimode fiber leads to a field with Gaussian statistics at the fiber output, equal to the Gaussian statistics of a TLS \cite{Mehringer2013}. By mechanically shaking the fiber the modes are dynamically mixed leading to the required variation of the pseudothermal field in time. Since multiphoton interferences of classical sources can be measured in the high-intensity regime \cite{Chekhova2008}, a conventional digital camera located in the far field of the fibers ($z \approx 0.40$~m) was used to measure the light intensity. Here, each pixel of the camera can be regarded as an individual detector. Intensity correlations of arbitrary order $g^{(m)}(\delta_1,\dots,\delta_m)$ can be derived by correlating the gray values of $m$ pixels at $\delta_j$, $j = 1, \ldots, m$ (see Fig.~\ref{fig:inv_setup}) \cite{Oppel2014}. A linear polarizer was placed in front of the camera to ensure that light of equal polarization was used.

One-dimensional arrangements of pseudo-TLS with varying sets of source separations $\mathbf{x}$  were realized by placing the end facets of the fibers onto grooves of a mechanical grid with lattice constant $d=570\,\mu$m. In this way the source geometries displayed in Fig.~\ref{fig:single_shots} have been implemented.
To obtain interference signals of high visibility, the integration time of the camera $\tau_i \sim 100$~$\mu$s was chosen to be much shorter than the coherence time of the TLS ($\tau_c \sim 10$~ms).

The experimental results for three different source arrangements are shown in Fig.~\ref{fig:single_shots}. For each setup we collected $N=1000$ camera images, each with a different realization of the pseudothermal field. The intensity distribution was confirmed to be thermal by measuring the instantaneous intensities at each pixel over the set of $1000$ camera images \cite{Mehringer2013}.
By correlating $m-1$ pixels at the magic positions (see Eq.~(\ref{eq:4})) with another pixel at $\delta_1$ we derived $g^{(m)}(\delta_1)$ for $m = 3, \ldots, 6$. Note that the finite lateral extension of the pseudothermal sources should prinicpally lead to a spatial envelope of $g^{(m)}(\delta_1)$. However, due to the small size of the fiber cores this modification is small and can be neglected (see Fig.~\ref{fig:single_shots}). This allows to use Eq.~(\ref{eq:5}) as a fit function for the experimental results. In this way we were able to determine from the modulations displayed in Fig.~\ref{fig:single_shots} the set of spatial frequencies $\textbf{F}$ from a least square fit.

According to the theory all occurring spatial frequencies $\textbf{F}$ are integer numbers (see Eq.~(\ref{eq:6})). This is excellently confirmed by the experimental results (see Table~\ref{tab:exp_results}), validating the outlined reconstruction algorithm. Moreover, applying the algorithm leads to a unique solution for all three investigated source arrangements as shown  on the left of Fig.~\ref{fig:single_shots}.

From Fig.~\ref{fig:single_shots} one can see that the measured amplitudes do not match the theory equally well as the spatial frequencies; moreover, they  show larger standard deviations (see Table~\ref{tab:exp_results}). This can be explained, among others, by the discrete size of the CCD pixels preventing the $m-1$ fixed detectors from being located exactly at the magic positions. Due to this inaccuracy sometimes a modulation can be seen in the $g^{(m)}$-signal although a constant is expected (see, e.g., $g^{(6)}_{3}(\delta_1)$ in Fig.~\ref{fig:single_shots}). However, trying to fit these signals with a modulated function leads to extraordinary large standard deviations of the fitted frequencies making these cases easily identifiable.

In conclusion we presented a new imaging protocol making use of linear optical detection techniques in combination with spatial intensity correlation functions of order $m \geq 3$ to derive the complete set of different spatial frequencies of an arbitrary irregular source arrangement in one dimension. The scheme allows to isolate the spatial frequencies of the system by use of the spatial intensity correlation functions $g^{(m)}$; in this way the relevant information about the source distribution can be extracted with a substantially reduced number of fit parameters in comparison to classical imaging techniques. 
Linking the set of different spatial frequencies $\mathbf{F}$ to the set of source distances $\mathbf{x}$ allows in most cases for a unique reconstruction, i.e., imaging, of the irregular source distribution. Remaining ambiguities can be removed by taking the amplitudes of the higher order intensity correlation functions into account. 
The scheme allows for subclassical imaging, i.e., it requires a numerical aperture smaller than the classical Abbe limit. 
Experimental results verifying the theoretical predictions were presented. 
As this approach is independent from the  photon wavelength and works without refractive optical elements potential applications in x-ray imaging, e.g., in astronomy, biology, medicine and the technical sciences, are expected.

The authors gratefully acknowledge funding by the Erlangen Graduate School in Advanced Optical Technologies (SAOT) by the German Research Foundation (DFG) in the framework of the German excellence initiative. D.B. gratefully acknowledges financial support by the Cusanuswerk, Bisch\"ofliche Studienf\"orderung.

\clearpage

\section*{Supplemental Material: Mathematical explanation of spatial frequency filtering}

For thermal light sources (TLS) the $m$th-order correlation function $g^{(m)}_{N\text{\,TLS}}(\delta_1,\ldots,\delta_m)$ contains $N^m$ multi-photon quantum paths, where each accumulates an individual phase. These phases have to be summed up coherently and incoherently according to their indistinguishable and distinguishable
\textit{final states}, respectively, and hence the correlation function results from considering all possible final states.
In the following we will show that by looking at the \textit{prefinal states} instead,
i.e., all detectors but the moving detector $D_1$ have already measured a single photon, we are able to derive a mathematical and physical explanation of the suppression of certain spatial frequencies.

When applying the magic positions to the detectors $D_{2},\ldots , D_{m}$ an astonishing effect appears, namely the suppression of all spatial frequencies, except for the frequency $(m-1)\delta_{1}$ and its higher harmonics in $g^{(m)}_{N\text{\,TLS}}(\delta_1,\ldots,\delta_m)$.
Considering the quantum path formalism the (not normalized) $m$th-order correlation function is given by the complex expression
\begin{align}
  \label{gN_m}
	 & G^{(m)}_{N \,\text{TLS}}(\delta_1,\ldots,\delta_m) \nonumber \\[2mm]
  &  = \langle\hat{E}^{(-)}(\delta_1)\ldots\hat{E}^{(-)}(\delta_m)\hat{E}^{(+)}(\delta_m)\ldots\hat{E}^{(+)}(\delta_1)\rangle_{\hat{\rho}} \nonumber \\[2mm]
	& = \sum_{\{n_{l}\}}X_{\{n_{l}\}}|\sum_{\mathcal{P}_{\{\alpha_{l}\}}}e^{i(\alpha_{l_{1}}\delta_{1}+\alpha_{l_{2}}\delta_{2}+\cdots+\alpha_{l_{m}}\delta_{m})}|^{2} \, ,
\end{align}
where the electric field operator
\begin{equation}
\hat{E}^{(+)}(\delta_j) = [\hat{E}^{(-)}(\delta_j)]^{\dagger} \propto \sum_{l=1}^{N} \hat{a}_l e^{i  \alpha_{l} \delta_{j}} \ ,
\end{equation} 
has been used, with the relative phase position $\delta_j = k d \sin(\theta_j )$ of the $j$th detector (see Fig.~1 in the main text) and the sources' relative phase prefactors  $\alpha_{l}$ (cf. Tab.~\ref{tab:N2}).

In Eq.~(\ref{gN_m}), $\sum_{\{n_l \}}$ sums over all possible $m$-photon distributions, i.e., final states, where $\{n_l \}=\{ n_1, \ldots , n_N \}$ describes the exact partitioning of the $m$ photons according to the emitting sources such that $\sum_{i=1}^{N}n_i = m$.
$X_{\{n_{l}\}}$ then denotes the statistical loading according to the normalized light statistics of the light field $\hat{\rho}$ and $\mathcal{P}_{\{\alpha_{l}\}}$ is the permutation over all phase prefactors $\alpha_l$ within a certain final state.

The phase prefactors $\alpha_l$ correspond to the relative distance between the first and the $l$th source in units of the lattice constant $d$ and are given in Tab.~\ref{tab:N2}. Note that the complete set of phase prefactors $\{\alpha_l \}=\{\alpha_1,\hdots,\alpha_1,\alpha_2,\hdots,\alpha_N,\hdots,\alpha_N\}$ contains $n_{l}$ times the prefactor $\alpha_l$ ($l=1,\hdots, N$), since the $l$th source has emitted $n_l$ photons. Note further that when looking at the prefinal states only $(m-1)$ photons are being emitted such that $\sum_{l=1}^{N}n'_l = m-1$.

\renewcommand{\arraystretch}{1.4} 
\begin{table} [ht]
\begin{center}
		\begin{tabular}{|c|ccccc|}
		\hline
			source number $l$ & \hspace{1mm} \ 1 \ & \ 2 \ &  \ 3 \ &\ $\hdots$ \ & \ N \\
			\hline
			phase prefactor $\alpha_{l}$ & \hspace{1mm} 0 & $x_1$ & $x_1+ x_2$ &  $\hdots$ &  $\sum_{l=1}^{N-1}{x_l}$ \\
			\hline
			$\#$ photons: prefinal state & \hspace{1mm} $n'_1$ & $n'_2$ & $n'_3$ &  $\hdots$ &  $n'_l$ \\
			\hline
			$\#$ photons: final state & \hspace{1mm} $n_1$ & $n_2$ & $n_3$ &  $\hdots$ &  $n_l$ \\
			\hline
\end{tabular}
\end{center}
\vspace{-3mm}
\caption{Source numbers with their corresponding relative phase prefactors.}
\label{tab:N2}
\end{table}
\renewcommand{\arraystretch}{1.0} 

Sorting the permutations of the phase prefactors $\{\alpha_{l}\}$ with respect to detector $D_{1}$, i.e., the last photon emission, one obtains (cf. Eq~(\ref{gN_m}))
\begin{align}
	&G^{(m)}_{N\,\text{TLS}}(\delta_{1},\hdots ,\delta_{m}) =  \sum_{\{n_{l}\}}X_{\{n_{l}\}} \nonumber\\
	& \times|\ c_{1}e^{i \alpha_1\delta_{1}}\sum_{\mathcal{P}_{\{\alpha_{l}\}\backslash \alpha_1}}e^{i(\alpha_{l_{2}}\delta_{2}+\alpha_{l_{3}}\delta_{3} +\cdots+\alpha_{l_{m}}\delta_{m})} \nonumber\\
	&\phantom{\times} + c_{2}e^{i \alpha_2\delta_{1}}\sum_{\mathcal{P}_{\{\alpha_{l}\}\backslash \alpha_2}}e^{i(\alpha_{l_{2}}\delta_{2}+\alpha_{l_{3}}\delta_{3}+\cdots+\alpha_{l_{m}}\delta_{m})} + \cdots  \nonumber\\
	&\phantom{\times} + c_{N}e^{i \alpha_N\delta_{1}}\!\!\!\sum_{\mathcal{P}_{\{\alpha_{l}\} \backslash \alpha_N}}\!\!\! e^{i(\alpha_{l_{2}}\delta_{2}+\alpha_{l_{3}}\delta_{3}+\cdots+\alpha_{l_{m}}\delta_{m})}\ |^{2} \ ,
\label{GN_m_2}
\end{align}
where $c_{l}=1$ ($c_{l}=0$) for $n_{l}>0$ ($n_{l}=0$). $\mathcal{P}_{\{\alpha_{l}\}\backslash \alpha_i }$ ($i=1,\hdots,N$) now describes all permutations of the set $\{\alpha_{l}\}$ with one phase prefactor $\alpha_{i}$ missing.

From Eq.~(\ref{GN_m_2}) it can be seen that each sum $\sum_{\mathcal{P}_{\{\alpha_{l}\}\backslash \alpha_i}}$ denotes all possible prefinal $(m-1)$-photon quantum paths that can lead to the final state $\{n_{l}\}$ with detector $D_{1}$ measuring a photon from the $i$th source, i.e., a specific prefinal state.


By interchanging the equivalent permutations $\mathcal{P}_{\{\alpha_l\} \backslash \alpha_i} \leftrightarrow \mathcal{P}_{\{\delta_j\}\backslash \delta_1}$, i.e., permuting the detectors instead of the sources, and rearranging the expressions in Eq.~(\ref{GN_m_2}) we obtain a sum of product terms for each prefinal state 
\begin{align}
\label{eq:factor}
&\sum_{\mathcal{P}_{\{\delta_{j}\}\backslash \delta_1}}e^{i(\alpha_{l_2}\delta_{j_{2}}+\alpha_{l_3}\delta_{j_{3}}+ \cdots +\alpha_{l_m}\delta_{j_{m}})}  \nonumber \\
	&\ =\sum_{j_{2}=2}^{m}e^{i\alpha_{l_2}\delta_{j_{2}}}\sum_{j_{3}=2}^{m}e^{i\alpha_{l_3}\delta_{j_{3}}} \cdots \sum_{j_{m}=2}^{m}e^{i\alpha_{l_m}\delta_{j_{m}}} \nonumber \\
	&\ - \Big[\sum_{j_{2}=2}^{m}e^{i(\alpha_{l_2}+\alpha_{l_3})\delta_{j_{2}}}\sum_{j_{3}=2}^{m}e^{i\alpha_{l_4}\delta_{j_{3}}} \cdots \sum_{j_{m}=2}^{m}e^{i\alpha_{l_m}\delta_{j_{m}}} \nonumber \\
	 &\ - \sum_{j=2}^{m}e^{i(\alpha_{l_2}+\alpha_{l_3}+\cdots+\alpha_{l_m})\delta_{j}}-\cdots\Big]\nonumber \\
	&\ - \Big[.\ .\ . \Big] - \cdots -\sum_{j=2}^{m}e^{i(\alpha_{l_2}+\alpha_{l_3}+\cdots+\alpha_{l_m})\delta_{j}} \, .
\end{align}
Each summand of Eq.~(\ref{eq:factor}) takes the form
\begin{equation}
\label{eq:factor2}
	\sum_{j_{2}=2}^{m}e^{i\alpha'_{l_2}\delta_{j_{2}}}\sum_{j_{3}=2}^{m}e^{i\alpha'_{l_3}\delta_{j_{3}}} \cdots \sum_{j_{k}=2}^{m}e^{i\alpha'_{l_k}\delta_{j_{k}}} \ ,
\end{equation}
where $k \leq m $ and the prefactors $\alpha'_{l_i}$ are consisting of combinations of the relative phase factors  $\alpha_{l_i}$. Each $\alpha_{l_i}$ has to be used and contributes exactly one time, leading to the surjective mapping $\alpha_{l_i} \rightarrow \alpha'_{l_i}$. Thus the equality $\sum_{i=2}^{k}\alpha'_{l_i}=\sum_{i=2}^{m}\alpha_{l_i}$ has to be fulfilled.

In the next step the magic positions
\begin{equation}
\delta_j = \frac{2\pi (j-2)}{(m-1)} \qquad j=2,\ldots,m \, ,
\label{eq:4a}
\end{equation}
are inserted and the relation
\begin{align}
\label{eq:mthroots1}
	\hspace{-5mm} \sum_{j=2}^{m}e^{i \lambda \delta_{j}}=\begin{cases}0 \ \ &, \lambda\neq\{0\} \ , \ \text{mod}(m-1)\\(m-1)  \ &, \lambda=\{0\} \ , \ \text{mod}(m-1) \, , \end{cases} \nonumber \\[-6.75mm] 
\end{align}
valid for the $(m-1)$th roots of unity, is applied.  As a result the summand of Eq.~(\ref{eq:factor2}) vanishes if at least one $\alpha'_{l_i} \neq \{0\} \ , \ \text{mod}(m-1)$. Hence all $\alpha'_{l_i} \stackrel{!}{=} \{0\} \ , \ \text{mod}(m-1)$ for a non-vanishing contribution and 
\begin{align}
\label{eq:factor4}
\sum_{i=2}^{m}\alpha_{l_i}= \sum_{i=2}^{k}\alpha'_{l_i}= \{0\} , \, \text{mod}(m-1) \ ,
\end{align}
has to be true. As the mapping $\alpha_{l_i} \rightarrow \alpha'_{l_i}$ is surjective for every summand of Eq.~(\ref{eq:factor}),
the entire prefinal state of Eq.~(\ref{eq:factor}) vanishes if $\sum_{i=2}^{m}\alpha_{l_i} \neq \{0\} , \, \text{mod}(m-1)$ and consequently will not contribute to
$g_{N\,\text{TLS}}^{(m)}(\delta_1,\ldots,\delta_{m})$.

Therefore, we can determine the contributing final states with the help of the valid prefinal states (cf. Eq.~(\ref{eq:factor4})). These can be denoted in the following form for a setup with $N$ sources:
\begin{align}
&\text{prefinal state:} \quad (n'_1,\hdots,n'_N)_{\text{pf}} \nonumber \\[0mm] &\qquad \text{with} \quad \sum_{l=1}^{N}{n'_l}= (m-1) \, ,
\label{eq:prefinal1}
\end{align}
where again $n'_l$ ($l=1,\ldots,N$) represents the number of photons emitted by the $l$th source. The relative phase factors accumulated for the $(m-1)$-photon detection event can be described by
\begin{align}
n'_1 &\, \alpha_1+ \cdots +  n'_N \, \alpha_N \nonumber \\[2mm]
& = \underbrace{(\alpha_1 + \cdots + \alpha_1)}_{n'_1\ \text{times}}  +\cdots + \underbrace{(\alpha_N + \cdots + \alpha_N)}_{n'_N\ \text{times}} \nonumber \\
& \stackrel{!}{=} \sum_{i=2}^{m}\alpha_{l_i}  \, ,
\label{eq:prefinal2}
\end{align}
where the $\alpha_{l_i}$ do not have to be pairwise different, since TLS can emit more than one photon. The contributing final states, arising from a particular prefinal state, are then given by:
\begin{align}
\label{eq:pre-fin1}
\hspace{-8mm} (n'_1,n'_2,\ldots,n'_N)_{\text{pf}} \rightarrow 
\begin{cases}
&  ((n'_1+1),n'_2,\ldots,n'_N)_{\text{f}}   \\
&		(n'_1,(n'_2+1),\ldots,n'_N)_{\text{f}}	\\
&  \; \vdots  \\
&		(n'_1,n'_2,\ldots,(n'_N+1))_{\text{f}}	 \, .
\end{cases} \nonumber \\[-5.85mm] 
\end{align}
Considering that the last photon is detected by the detector $D_1$ the accumulated relative phase prefactors can be calculated to (cf. Eq.~(\ref{eq:factor4}))
\begin{align}
\label{eq:finPhase}
\sum_{i=1}^{m}\alpha_{l_i} = \sum_{i=2}^{m}\alpha_{l_i} \,+\; \alpha_{l_1} = \{0\} ,  \text{mod}(m-1) + \; \alpha_{l_1} \,.
\end{align}
To obtain interference, due to coherent summation of accumulated phases, at least two \textbf{different} prefinal states yielding the \textbf{same} final state are necessary. This leads to the condition
\begin{align}
 \; \left(\alpha_{l_1} \right)_{\text{f1}}- \left(\alpha_{l_1} \right)_{\text{f2}} &=  \left(\sum_{i=2}^{m}\alpha_{l_i}\right)_{\text{pf2}} - \left(\sum_{i=2}^{m}\alpha_{l_i}\right)_{\text{pf1}} \nonumber \\[2mm]
 &= \{0\} , \, \text{mod}(m-1) \ ,
\label{eq:fin-two-phase2}
\end{align}
for the last photon detections at detector $D_1$. This result is equivalent to the phenomenological observation of Eq.~(4) in the main text. Interference can only be produced from pairs of sources $R_{l_1}$ and $R_{l_2}$ whose difference in phase prefactors, i.e., whose separation $(x_{l_1}+\cdots+x_{l_2-1})$ corresponds to $\{0\} , \, \text{mod}(m-1)$, whereas photons originating from other sources lead to a constant offset.

\end{document}